\documentclass[paper,11pt]{JHEP}
\pdfoutput=1
\usepackage{graphicx}
\usepackage[centertags]{amsmath}
\usepackage{amsfonts} 
\usepackage{amssymb} \usepackage{amsthm}
\usepackage{multirow}
\usepackage{wasysym}
\usepackage{bbold}
\allowdisplaybreaks[1]
\def\beq{\begin{equation}}
\def\eeq{\end{equation}}

\newcommand{\eqa}{\begin{eqnarray}}
\newcommand{\ena}{\end{eqnarray}}
\newcommand{\eq}[1]{eq. (\ref{#1})}

\newcommand{\Z}{\mathbb{Z}}
\newcommand{\N}{\mathbb{N}}

\newcommand{\OO}{{\cal O}}

\newcommand{\bra}{\langle}
\newcommand{\ket}{\rangle}

\title{Critical exponents of the 3d Ising and related models from Conformal Bootstrap }
\author{
Ferdinando Gliozzi$^{a,b}$, Antonio Rago $^a$\\
$^a$ School of Computing and Mathematics and Centre for Mathematical Science, Plymouth University, Plymouth PL4 8AA, United Kingdom\\
$^b$ Dipartimento di Fisica, Universit\`a di Torino\\
and Istituto Nazionale di Fisica Nucleare - sezione di Torino \\
Via P. Giuria 1, I-10125 Torino, Italy

}
\bigskip
\abstract{
The constraints of conformal bootstrap are applied to investigate a set of conformal field theories in various dimensions. The prescriptions can be applied to both
unitary and non unitary theories allowing for the study of the
spectrum of low-lying primary operators of
the theory. We evaluate the lowest scaling dimensions of
the local operators associated with the Yang-Lee edge singularity for $2\le D\le6$. Likewise we obtain the scaling dimensions of
six scalars and  four spinning  operators for the 3d critical Ising
model. Our findings are in agreement with existing results to a per mill precision
and estimate several new exponents.
}

\begin{document}
\section{Introduction}

Understanding conformal field theories (CFTs) is a great challenge of many branches in theoretical physics. CFTs are at the heart of critical phenomena 
in condensed matter physics, they control the renormalisation group flows of quantum field theories and explain the appearance of universal scaling 
laws \cite{Polyakov:1970xd,Wilson:1973jj}; they even provide a tool to
study quantum gravity via the AdS/CFT correspondence \cite{Maldacena:1997re}. 

One of the main properties characterizing a specific CFT is the spectrum of the 
scaling dimensions of its local operators. 
With the exception of some two-dimensional CFTs (where the  algebra of conformal generators is infinite-dimensional) calculating these quantities 
is very challenging, as they are dominated by  quantum fluctuations,
an effect which takes place on all length scales in these
scale-invariant theories.  Moreover, most CFTs are strongly coupled and difficult 
to study using the usual  perturbative techniques of Feynman diagrams, although some of them can be accurately analysed by Monte Carlo calculations and/or strong coupling expansions \cite{Deng:2003wv,phi4pisa,Martin2010}.

Conformal bootstrap is a reincarnation of the bootstrap 
mechanism used to investigate the  CFT constraints originating from the crossing
symmetry of the four-point functions. It was suggested a long time ago that conformal bootstrap could give useful informations on the allowed scaling dimensions of the theory \cite{{Ferrara:1973yt},{Polyakov:1974gs},
{Polyakov:1984yq}}. This idea was implemented in two-dimensional rational CFTs, i.e. those with a finite number of Virasoro primary fields. It was shown that the crossing symmetry, when combined with the modular invariance of the theory on a torus, provides the complete spectrum of scaling dimensions, modulo an integer
 \cite{Vafa:1988ag}. 

In recent years it has been shown that the conformal bootstrap
approach can give accurate predictions for specific CFTs in any space-time dimension
\cite{Rattazzi:2008pe,Rychkov:2009ij,Rattazzi:2010gj,Poland:2010wg, ElShowk:2012ht,Pappadopulo:2012jk,Liendo:2012hy, ElShowk:2012hu,Komargodski:2012ek,Gliozzi:2013ysa,Kos:2013tga,El-Showk:2013nia,
Gaiotto:2013nva}. The starting point of this active field of research was 
the  observation \cite{Rattazzi:2008pe}  that the conformal bootstrap
equations (i.e the functional equations following from the crossing symmetry) can be rewritten 
as an infinite system of linear homogeneous equations.
 In order to study the system numerically the number of unknowns and
 the number of equations has to be truncated. 
In the simplest case, i.e. the four-point function of a single scalar field $\phi$ with scaling dimension $\Delta_\phi$, the truncated  bootstrap equations
take the form
\beq
\sum_{i=1}^N{\sf f}^i_e\,{\sf p}_i=0\,,\;(e=1,2,\dots,M)\,,
\label{homo}
\eeq   
where the unknown ${\sf p}_{i}$ is the square of the coefficient of the primary
operator $\OO_i$ contributing to the operator 
product expansion (OPE) of $\phi(x)\phi(y)$. 
The coefficient ${\sf f}^i_e$ -- made up of multiple derivatives of generalized hypergeometric functions -- depends on $\Delta_\phi$ and on the scaling dimension 
$\Delta_i$ of $\OO_i$. 
Unitarity requires ${\sf p}_i\ge0$; this 
implies that the search of solutions for \eq{homo}, when 
combined with a normalization condition, can be reformulated as a
linear programming problem, which can be treated numerically. 
Its solution  yields a numerical upper bound $\Delta_u=f(\Delta_\phi)$ on the dimension of the first 
scalar  contributing to the OPE of  $\phi(x)\phi(y)$
\cite{Rattazzi:2008pe}. 
Some CFTs may exist that saturate this bound\cite{ElShowk:2012ht,Kos:2013tga}.
Since the output of the linear programming algorithm is a solution 
of \eq{homo}, 
the low-lying spectrum of operator dimensions for these special
CFTs can be evaluated. The accuracy of these calculations is
particularly impressive in the case of the two-dimensional critical
Ising model, where a comparison with exact results can be made.
A drawback of this approach is that it can be applied only to theories 
saturating the unitarity bound. 
In addition, a physically convincing reason why some CFTs should saturate the bound is still missing\footnote{In section \ref{crossing}  the condition for an actual CFT to saturate this unitarity bound is rewritten in a closed form.}.      

Recently one of us has proposed a different approach to solve numerically 
the bootstrap constraints (\ref{homo}) which can be applied to a
larger class of CFTs \cite{Gliozzi:2013ysa}. 
The starting point was the observation that 
any finite truncation of the bootstrap equations results in a set constraints
on the spectrum of operator dimensions contributing to \eq{homo},
provided that the number $M$ of equations is equal or larger than the number
$N$ of unknowns. In this case the homogeneous system admits a
non-identically vanishing solution if and only if all the minors of
order $N$ are vanishing. 
Hence for any subset of $N$ equations the constraint can be expressed as
\beq
 \det\,{\sf f_n}\equiv f_n(\Delta_\phi,\Delta_1,\Delta_2,\dots,\Delta_N)=0\,,
~~(n=1,2,\dots,\left(\begin{matrix}M\cr N\cr\end{matrix}\right))
\label{det}
\eeq   
where ${\sf f_n}$ is the $N\times N$ matrix of  coefficients of the
$n$th subset. As soon as $M$ exceeds $N$, the number of constraints is
equal or larger than $N$; the all set of constraints can be used to extract the scaling dimensions of the primary
operators involved, if a solution exists. If the system of constraints is over-determined, i.e. there are more
 independent equations than unknowns, it can be split in consistent
 subsystems to obtain a set of solutions; their spread provides a
 rough estimate of the error. 
A caveat: these solutions are intrinsically affected by an unknown systematic
error due to the truncation in the number $N$ of operators. Such
systematic error is bound to decrease when the number of included operators increases.

This method is quite general and it can be applied to any CFT, whether
unitary or not, provided the theory has a discrete spectrum and it admits a consistent 
truncation (\ref{homo}) of the bootstrap equations.
This statement hides a subtlety that deserves to be mentioned. 

The number $M$ of homogeneous equations is not an 
independent parameter: indeed the larger is the number $N$ of terms in 
(\ref{homo}), the better is the approximation of the rhs to a constant, and 
hence the larger is the number $M$ of approximately vanishing derivatives.
 
The method outlined above is valid only if it exists an $N_0$ such that $M>N_0$.
If this happens for small values of $N_0$ we say that the CFT is
 easily truncable. 
Typical examples  are the free scalar massless theories in $D$ dimensions, where the method yields very accurate results already 
for $N_0=3$  \cite{Gliozzi:2013ysa}.
In spite of the fact that the conformal block expansion shows good convergence 
properties \cite{Pappadopulo:2012jk}, there are cases in which the assumption
$M>N$ can not be consistently made  for any value of $N$ 
(an explicit example is shown in the next section). In these cases 
the associated CFT is not truncable and no numerical method based
  on the analysis of just one four point function can be applied.

Of course the method is predictive only if
the exact form of the four-point function for a given CFT is not known.
Should this be the case, we do not know {\sl a priori} whether the theory 
is easily truncable. 
The symmetry properties of the model can be used to guess the 
structure of the quantum numbers of the low-lying primary operators.
If a valid set of zeros for $N$ determinants of the type  (\ref{det}) is found, we can say {\sl a 
posteriori}  that the CFT is easily truncable. 
The larger is $N$ the smaller is the error of the estimate of the
operator dimensions. Thus, once a solution is found, it can be improved by adding new operators in the game.
In this way the low-lying spectrum of the Yang-Lee model in several dimensions as been evaluated in \cite{Gliozzi:2013ysa} and will be further improved in this paper.

A major result presented in this paper is an exact solution of the conformal bootstrap constraints (\ref{det}) with $N=7$ primary operators and $M=8$ equations.
It represents a consistent truncation of the four-point function 
$\bra\sigma(x_1)\sigma(x_2)\sigma(x_3)\sigma(x_4)\ket$ of the 3d Ising model at 
the critical point described by the spontaneous breakdown of the $\Z_2$ symmetry. 
$\sigma(x)$ is the $\Z_2$-odd scalar, i.e. the order parameter of the theory. 
The set of primary operators comprises: four $\Z_2$-even scalars, a
conserved spin 2 operator, a quasi-conserved spin 4 operator and a spin 6 operator. 
One of the four $\Z_2$-even scalars is identified with 
the energy $\varepsilon$, the only relevant operator beside $\sigma$, 
while the other three are recurrences, $\varepsilon',\varepsilon'',\varepsilon'''$, whose dimensions 
are related to the critical exponents measuring the corrections to
scaling.
The spin 2 operator is identified with the stress tensor, while the
spin 4 operator is related to the critical
exponent associated with the rotational symmetry breaking.
It turns out that the solution depends on a free parameter. Using the scaling dimension
of the spin 4 operator as an input all the other  operator dimensions
can be evaluated and they differ from the best estimates for no more
than 1 $\permil$. 

Assuming  this truncation is not only a solution of  
crossing symmetry constraints,  but a part of a full-fledged CFT, 
we exploit some consistency checks to further enlarge the number of estimated  
scaling dimensions of  primary operators belonging to the 
spectrum of the 3d critical Ising model, as reported in Tables \ref{tab:4} and \ref{tab:5}.

Even though a formal proof of conformality of the critical point of
the 3d Ising model is still missing \cite{Polchinski:1987dy}, numerical simulations and theoretical arguments leave 
little doubt on its validity.  The conformal bootstrap approach
  \cite{ElShowk:2012ht} strongly supports this hypothesis. This paper, whose results are based
  solely on the assumption of conformal invariance of the 3d Ising
  critical point, further corroborates the hypothesis.

The plan of the paper is the following. In the next Section we fix the notations  and  discuss the derivation of the homogeneous system (\ref{homo}) 
in relation with the notion of easily truncable CFT. Section
\ref{YangLee} is dedicated to Yang-Lee models in various dimensions
while the last Section is devoted to the 3d critical Ising model and to some conclusions.   
\section{Conformal Bootstrap Constraints}
\label{crossing}
The four-point function of a scalar $\phi(x)$ of a $D$-dimensional CFT can be parametrised as \cite{Polyakov:1970xd}     
\beq
\bra\phi(x_1)\phi(x_2)\phi(x_3)\phi(x_4)\ket=
\frac{g(u,v)}{\vert x_{12}\vert^{2\Delta_\phi}\vert x_{34}\vert^{2\Delta_\phi}},
\label{fourpoint}
\eeq 
where $\Delta_\phi$ is the scaling dimension of $\phi$,  $x_{ij}^2$ is the square of the distance between $x_i$ 
and $x_j$, $g(u,v)$ is a function of the cross-ratios 
$u=\frac{x_{12}^2x_{34}^2 }{x_{13}^2x_{24}^2 }$ and
 $v=\frac{x_{14}^2x_{23}^2 }{x_{13}^2x_{24}^2 }$.  

\noindent $g(u,v)$ can be expanded in terms of the conformal blocks $G_{\Delta,L}(u,v)$, i.e.  the eigenfunctions of the Casimir operator of $SO(D+1,1)$:
\beq
g(u,v)=1+\sum_{\Delta,L}{\sf p}_{\Delta,L}G_{\Delta,L}(u,v),
\label{bexpansion}
\eeq
with ${\sf p}_{\Delta,L}=\lambda_{\phi\phi\OO}^2$, where 
$\lambda_{\phi\phi\OO}$ is the coefficient of the primary operator 
$\OO$ of scaling dimension $\Delta$ and spin $L$ contributing to the OPE of $\phi(x)\phi(y)$. 

The lhs of (\ref{fourpoint}) is invariant under any permutation of the $x_i$'s while the rhs is not, unless $g(u,v)$ obeys the following two functional equations
\beq
g(u,v)=g(u/v,1/v)\,; \,\,v^{\Delta_\phi}g(u,v)=u^{\Delta_\phi}g(v,u)\,.
\label{fuequ}
\eeq
Since $G_{\Delta,L}(u,v)=(-1)^LG_{\Delta,L}(u/v,1/v)$, the first equation projects out the odd spins in the expansion (\ref{bexpansion}). 
The second one, after separating the identity from the other primary operators, can be rewritten as a sum rule
\beq
\sum_{\Delta,L}{\sf p}_{\Delta,L}
\frac{v^{\Delta_\phi}G_{\Delta,L}(u,v)-u^{\Delta_\phi}G_{\Delta,L}(v,u)}
{u^{\Delta_\phi}- v^{\Delta_\phi}}=1\,.
\label{sumrule}
\eeq

 It is easy to argue that such a relation  can be exactly satisfied only if the sum contains infinite terms \cite{Rattazzi:2008pe}. 
In any finite truncation  the rhs of (\ref{sumrule}) is only approximately constant. In order to transform the sum rule into the system (\ref{homo}) of 
linear homogeneous 
equations one has to assume that the derivatives of the 
rhs are vanishing, an assumption that in some cases 
cannot be made. 

As an example, consider the four-point function of the energy
operator $:\phi^2(x):=\lim_{\epsilon\to0}\phi(x)^i\phi_i(x+\epsilon)-
\bra\phi^j(x)\phi_j(x+\epsilon)\ket$ in a four-dimensional $O(n)$-invariant 
 free field theory $(i,j=1,2\dots,n)$.

\begin{figure}
\centering
\includegraphics[width=9 cm]{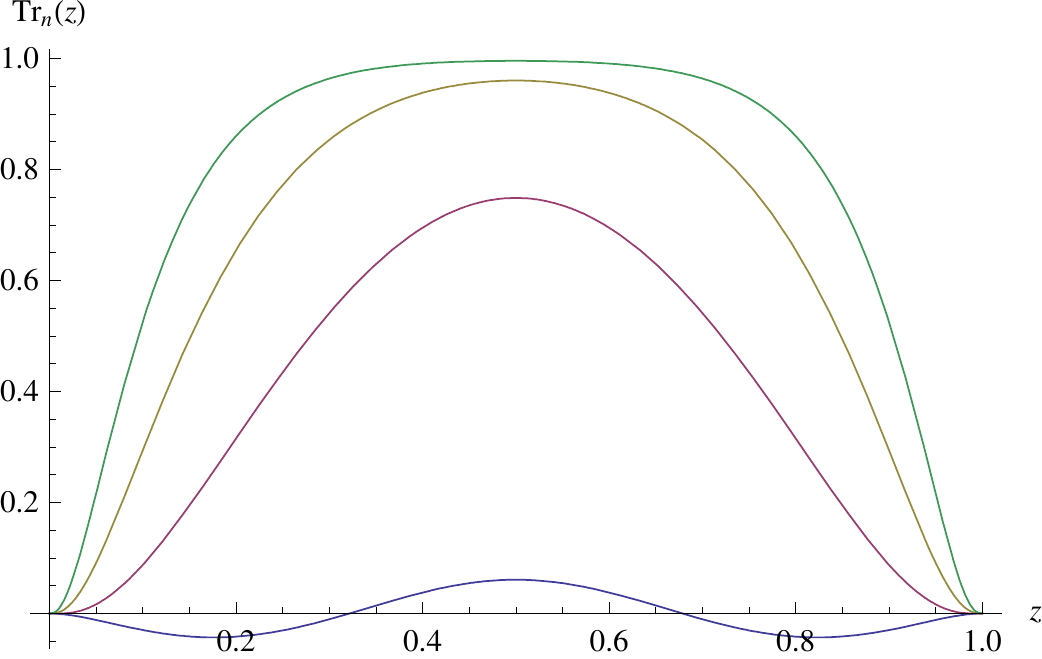}
\caption{ Progressive truncations $Tr_n(z)$ of the sum rule 
(\ref{sumrule}) at $z=\bar{z}$ for the 4-point function  of the
energy operator in the $4d$ $O(n)$-invariant  free massless scalar theory.
The four curves represent respectively the sum of the contributions of
the primaries of 
scaling dimensions $\Delta\le4,\,\Delta\le6,\,\Delta\le8,\,\Delta\le10$. In this specific example we have chosen $n=1$.}
\label{fig:1}
\end{figure}

We have
\beq
g_n(u,v)=1+\frac4n\left(u+\frac uv+\frac{u^2}v\right)+\frac{u^2}{v^2}+u^2.
\eeq 
The first few terms of the  conformal block expansion are
\beq
g_n(u,v)=1+\frac8n G_{2,0}+(2+\frac4n)G_{4,0}+\frac{16}{3n}G_{4,2}+
(2-\frac2n)G_{6,0}+\frac{24+\frac8n}{5}G_{6,2}+\frac{64}{35n} G_{6,4}+\dots
\label{gn}
\eeq
with the  parametrisation $u=z\bar{z}$ and
   $v=(1-z)(1-\bar{z})$ \cite{Dolan:2003hv}  which simplifies  the functional form of the conformal
   blocks.
 In the Euclidean space $z$ and $\bar{z}$ are complex conjugates of each
 other. 
Choosing  $z=\bar{z}$ corresponds to configurations with all four
points on a circle.  
Figure \ref{fig:1} shows  the  first 
few truncations. 
It should be clear that these latter solutions are not  
flat enough around the symmetric point $z=\bar{z}=\frac12$. 

In this case it would be erroneous to guess that higher level truncations could always generate a   number $M$ of homogeneous equations sufficient to solve the system. 
Actually this four-point function is not truncable. 
The proof is surprisingly simple: if $M$ is larger than $N$, then the knowledge of the operator spectrum should determine  the 
 ${\sf p}_{\Delta,L}$'s, but these depend on $n$ (see \eq{gn}), 
while the spectrum of primaries of such a free field theory does not depend on it. 
%NN%

\begin{figure}
\centering
\includegraphics[width=9 cm]{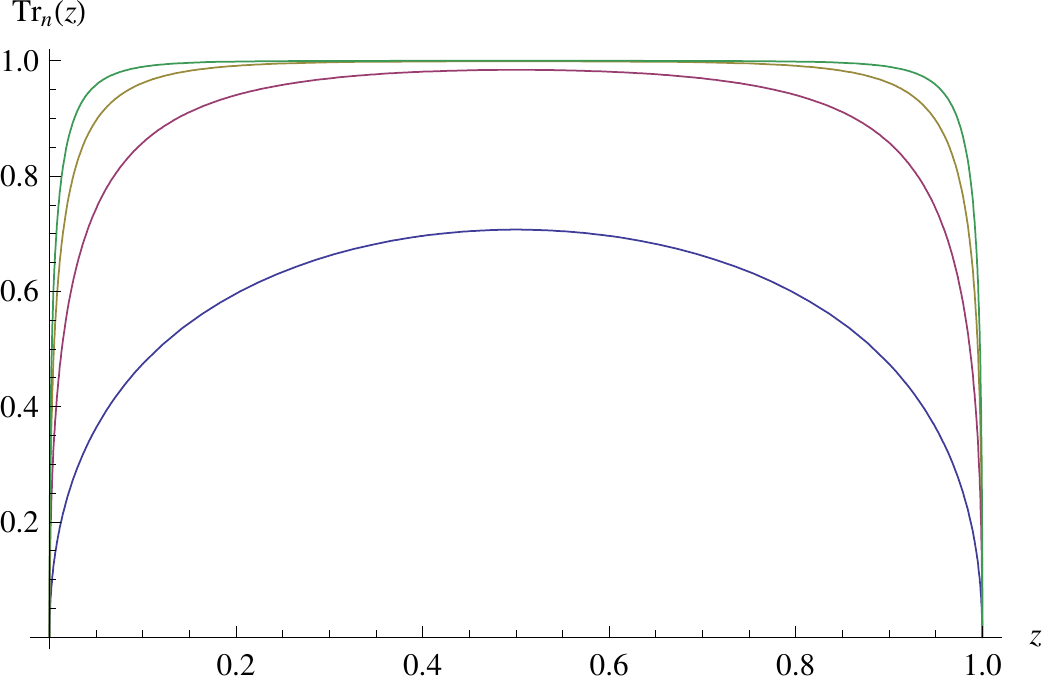}
\caption{ Progressive truncations $Tr_n(z)$  of the sum rule 
(\ref{sumrule}) for $z=\bar{z}$ of the 4-point function 
of a free massless scalar theory in $3d$. Few terms are sufficient to well approximate the rhs of the sum rule around the symmetric point $z=\frac12$.  }
\label{fig:2}
\end{figure}

The easily truncable CFTs are instead those where the first few terms suffice 
to give an almost constant rhs in (\ref{sumrule}). 
This is the case for instance of free scalar massless theories in $D$ dimensions. Figure \ref{fig:2} shows the first few truncations at $D=3$, where
\beq
g(u,v)-1\equiv \sqrt{u}+\sqrt{\frac uv} =2G_{1,0}+\frac14G_{3,2}+
\frac1{64}G_{5,4}+\frac1{1024}G_{7,6}+\dots
\label{free3} 
\eeq
In these cases one can follow the 
procedure of  \cite{Rattazzi:2008pe}, i.e.  Taylor expand (\ref{sumrule}) 
about the symmetric point $z=\bar{z}=\frac12$ and transform a $N$-terms truncation of the sum rule into  a set of $M$ linear equations in 
$N$ unknowns ${\sf p}_{\Delta,L}$. 

To be more specific, following  \cite{ElShowk:2012ht} we make the change of variables 
$z=(a+\sqrt{b})/2$, $\bar{z}=(a-\sqrt{b})/2$
and Taylor expand around $a=1$ and $b=0$.
 It is easy to see that this expansion will contain only even powers 
of $(a-1)$ and integer powers of $b$.
The truncated sum rule can then be rewritten as one inhomogeneous equation
\beq
\sum_{\Delta,L}{\sf p}_{\Delta,L}{\sf f}^{(0,0)}_{\Delta_\phi,\Delta_L}  =1,
\label{inh}
\eeq
and a set of $M$ homogeneous equations already mentioned in \eq{homo} and here rewritten in more details
 \beq
\sum_{\Delta,L}{\sf f}^{(2m,n)}_{\Delta_\phi,\Delta_L}{\sf p}_{\Delta,L} =0,~
(m,n\in\N,m+n\not=0),
\label{homos}
\eeq
with
\beq
{\sf f}^{(m,n)}_{\alpha,\beta}=\left(\partial_a^{m}\partial_b^n
\frac{v^{\alpha}G_{\beta}(u,v)-u^{\alpha}G_{\beta}(v,u)}
{u^{\alpha}- v^{\alpha}}\right)_{a=1,b=0}.
\label{matrix}
\eeq
The number $M$ of homogeneous equations depends on the degree of
flatness of the truncation. When $M>N$ the system (\ref{homos})
becomes highly predictive, since it admits a non identically vanishing
solution if and only if all the minors of order $N$ are vanishing. The
common intersection of these zeros identify the spectrum values of the 
scaling dimensions for the $N$ primary operators. If there are more independent minors than unknowns we get a set of scattered solutions. Their spread gives a rough estimate of the error of the approximation. Inserting these $\Delta$'s in 
\eq{homos} and assuming that the corresponding minor has rank $N-1$ (i.e. a simple zero) we obtain a one-parameter family of ${\sf p}_{\Delta,L}$'s. The inhomogeneous equation (\ref{inh}) sets their normalization.

\begin{figure}
\centering
\includegraphics[width=9 cm]{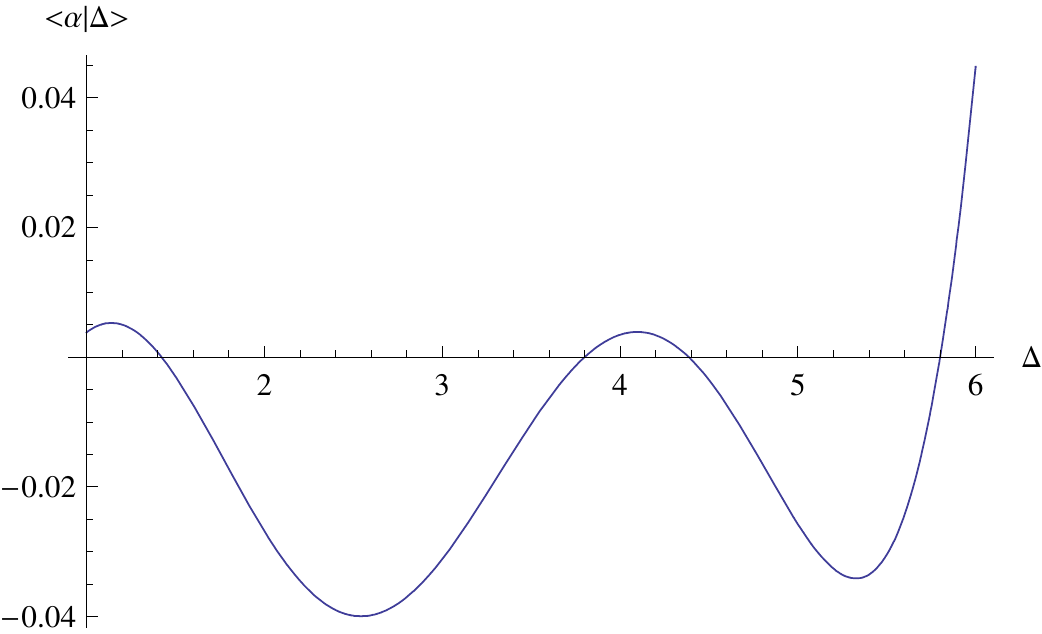}
\caption{The scalar product $\bra\alpha\vert\Delta\ket\equiv
\bra\alpha\vert 0,\Delta,\Delta_\phi\ket$  as a 
function of $\Delta$ 
in the critical 3d Ising model. $\bra\alpha \vert$ is the left eigenvector corresponding to the null eigenvalue of the $7\times7$ matrix ${\sf f}$ discussed in Section \ref{Ising} and $\vert\Delta\ket$ is a column of ${\sf f}$ associated with  conformal blocks of 
spin 0. The zeros  are the scaling dimensions of the 
$\Z_2$-even scalars $\varepsilon,\varepsilon',\varepsilon'',\varepsilon'''$.}
\label{fig:3}
\end{figure}

It is instructive to consider at the homogeneous system (\ref{homos})
from a slightly different angle. The  ${\sf p}_{\Delta,L}$'s can be
viewed as the components of the right-eigenvector with zero eigenvalue
of ${\sf f}$. In Dirac notation we can write ${\sf f} \vert {\sf
  p}\ket=0$. Of course there is also a left-eigenvector associated
with the same eigenvalue, i.e. $\bra\alpha\vert{\sf f}=0$.
It can be noticed that $\bra\alpha\vert$  encodes the complete
information on the low-lying spectrum of scaling dimensions of the
theory. Indeed $\bra\alpha\vert$ is by construction orthogonal to the
$N$  columns of ${\sf f}$. 
Each column depends only on a single conformal block, so it can 
be viewed as a vector that we denote as $\vert L ,\Delta,\Delta_\phi\ket$. 
By the knowledge of $\bra\alpha\vert$ it is possible to reconstruct
the whole spectrum of the $N$ primary operators as follows. Since
 the stress-tensor is always present among these operators, the zeros of $f(x)\equiv\bra\alpha\vert 2,D,x\ket$ give
the possible values of $\Delta_\phi$. It follows that the zeros of the scalar
product $\bra \alpha\vert  L ,\Delta,\Delta_\phi\ket$, as a
function of $\Delta$ and for fixed $L$ and $\Delta_\phi$, provide the whole
spectrum (see an example in figure \ref{fig:3}). In conclusion, the
combination of the
two eigenvectors $\vert{\sf p}\ket$ and $\bra\alpha \vert$ 
encodes the complete information to reconstruct the (truncated)
four-point function for the CFT under study.

  The eigenvector $\bra\alpha \vert$ bears some similarity to the 
linear functional $\bra\Lambda\vert$ discussed in \cite{Rattazzi:2008pe} in relation with the unitarity upper bound. In both cases the zeros of the scalar product with  $\vert L ,\Delta,\Delta_\phi\ket$
provide the operator spectrum 
\cite{Poland:2010wg,ElShowk:2012hu}, hence a solution to the crossing 
symmetry; on the other hand $\bra\Lambda\vert$ must fulfil the much stronger constraint
\beq
\bra\Lambda\vert L ,\Delta,\Delta_\phi\ket\ge 0\;,~~ \Delta\ge\Delta_L^o\,. 
\label{Lambda}
\eeq 
$\Delta_L^o$ is the upper unitarity bound described in the
introduction: if at a given  spin $L$ the first primary operator
$\OO_{\Delta,L}$ has scaling dimension larger than this bound, the
theory is not unitary. Thus if a unitary  CFT saturates the bound, the
scalar product above must vanish at $\Delta=\Delta_L^o$ and all the recurrences at the same $L$ must correspond to double zeros of 
(\ref{Lambda}), i.e.
\beq
\bra\Lambda\vert L ,\Delta,\Delta_\phi\ket=0,\;
\bra\Lambda\vert\frac{d}{d\Delta}\vert L ,\Delta,\Delta_\phi\ket=0,\;
\Delta\in\Sigma,\Delta>\Delta_L^o,
\eeq
where $\Sigma$ denotes the spectrum of allowed operator dimensions.

It follows that $\bra\Lambda\vert$ can be seen as a left eigenvector of  zero eigenvalue for a matrix ${\sf F}$ much larger than ${\sf f}$, composed by the columns $\vert L ,\Delta,\Delta_\phi\ket$ for all $\Delta\in\Sigma$ and the columns
$\frac{d}{d\Delta}\vert L ,\Delta,\Delta_\phi\ket$ for all  recurrences. 
For instance, if the set of $N$ operators of a given truncation of a CFT is composed by $n_0$ scalars, $n_2$ operators of spin 2 and so on, so that 
$N=n_0+n_2+\dots+n_{2k}$, the matrix ${\sf F}$ has $2n_0-1$ columns for the scalars, $2n_2-1$ columns for the spin 2, and so on, so that the total  number of columns is $N_F=2N-k$. 

Providing  a solution to the crossing symmetry is 
equivalent to finding a $N$-dimensional left eigenvector 
$\bra\alpha\vert\,{\sf f}=0$. The additional constraint of saturating 
the unitarity 
bound requires instead to find a solution of $\bra\Lambda\vert{\sf F}=0$. 
If $\bra\Lambda\vert$ is unique as claimed in \cite{ElShowk:2012hu}, then ${\sf F}$ is a square $N_F\times N_F$ matrix with $\det \,{\sf F}=0$ and rank $N_F-1$.

In this paper we only study the zeros of $\det {\sf f}$. An essential ingredient to pursuing high accuracy in these calculations is an efficient method to evaluate with high precision the conformal blocks and their derivatives. These are known in a closed form only for $D=2,4,6$ \cite{Dolan:2003hv}. The conformal blocks for generic $D$ can be still rewritten in a closed form 
for $z=\bar{z}$, $L=0$ and $L=1$ in terms of $_3F_2$ hypergeometric functions
\beq
G_{\Delta,0}(z)=\left(\frac{z^2}{1-z}\right)^{\Delta/2}\,
_3F_2\left[\frac\Delta2, \frac\Delta2,\frac\Delta2-\alpha;\frac{\Delta+1}2,\Delta-\alpha;\frac{z^2}{4(z-1)}\right],
\eeq
\beq
G_{\Delta,1}(z)=\left(\frac{z^2}{1-z}\right)^{\frac{\Delta+1}2}\,
_3F_2\left[\frac{\Delta+1}2,\frac{\Delta+1}2,\frac{\Delta+1}2-\alpha;
\frac\Delta2+1,\Delta-\alpha;\frac{z^2}{4(z-1)}\right],
\eeq
with $\alpha=\frac D2-1$.

 We follow the algorithms developed in \cite{ElShowk:2012ht} which allow 
to write, through the use of a few recursion relations, each  matrix coefficient of (\ref{homos}) as
\beq
{\sf f}^{(2m,n)}_{\Delta_\phi,\Delta_L}=\sum_{i=1}^6R_i(D,L,\Delta_L,\Delta_\phi)\, 
{\cal B}_i(D,\Delta_L,z=\frac12),
\eeq 
 where the six basis functions ${\cal B}_i$ are the two conformal blocks 
 $G_{\Delta,0}(z)$ and $G_{\Delta,1}(z)$ and their first and second derivatives with respect to $z$; the $R_i$'s are rational functions of their arguments.

These formulas work for any value of dimension, even fractional ones. As such they have already been used to study conformal bootstrap in non-integer 
dimensions  \cite{El-Showk:2013nia}. Here we use this property when
comparing our results with the epsilon expansion of the Yang-Lee model 
in $6-\epsilon$ dimensions.  
 
\section{The Yang-Lee model}
\label{YangLee}
The CFTs described in the present and in the next Section are
  related to the Wilson-Fisher point of a massless
  scalar field model perturbed by an interacting term of the form
  $\phi^n$, with $n=3,4$. 
In these theories the only difference between the operator content of
the interacting model and that of the free field theory (the Gaussian
fixed point) is in the redundant operators. 
Standard renormalisation group arguments are used to describe how the
fusion algebra of the free theory gets modified by approaching the
non-trivial fixed point ( see e.g. \cite{Cardy}).

Besides the CFT describing its critical point, there are two other CFTs, at least, related to the 3d Ising model. One is the one-dimensional CFT associated with a line defect of this model. The low-lying spectrum of scaling dimensions 
of the local operators living on the defect has been estimated using  Monte Carlo simulations only recently \cite{Billo:2013jda}; these estimates have been supported by both epsilon expansion and conformal bootstrap calculations \cite{Gaiotto:2013nva}. 

The other CFT is much older and originates from two seminal papers of Yang 
and Lee \cite{Yang:1952be,Lee:1952ig}, dated more than sixty years ago, on the analytic properties of phase transitions. It turns out that the zeros of the partition function of a ferromagnetic Ising model in $D$ dimensions in a magnetic field 
$h$ are located on the imaginary axis above a critical value $h_c(T)$
called the Yang-Lee edge singularity. Above the critical temperature $T>T_c$, in the 
thermodynamic limit, the density of these zeros behaves near 
$h_c$ like $(h-h_c)^\sigma$, where the edge exponent $\sigma$ is universal and characterizes a universality class which is not related to the spontaneous 
breakdown of any symmetry. The Yang-Lee universality class can be 
described by the non-trivial fixed point of a $\phi^3$ theory \cite{Fisher:1978pf} with imaginary coupling, thus the corresponding CFT is non-unitary hence it  
cannot be studied with the conformal bootstrap approach of 
\cite{Rattazzi:2008pe}. 

The critical exponent $\sigma$ is related to the scaling dimension of $\phi$ by
\beq
\sigma(D)=\frac{\Delta_\phi}{D-\Delta_\phi}~.
\eeq
The interest on the Yang-Lee universality class is enhanced by the discovery, in the past years, that the edge singularity is related to 
other quite different critical behaviours. For instance, the number per 
site of large isotropic branched polymers in a good solvent 
(i.e. {\sl lattice animals}) obeys a power law with  exponent 
$\varphi_I(D)=\sigma(D-2)+2$ \cite{Parisi:1980ia}, or the pressure of $D$-dimensional fluids with repulsive core  has a singularity at negative values of activity with universal exponent $\varphi(D)=\sigma(D)+1$ \cite{Park:1999}. 

Monte Carlo simulations and other numerical methods on these systems gave 
accurate results for $\sigma$ \cite{Lai:1995,Hsu:2005}. 
Recent high temperature expansions of the Ising model in low magnetic field  
have further improved the accuracy in the whole range 
$2<D<6$ \cite{Butera:2012tq}.

The edge exponent $\sigma$ is exactly known in $D=2$ and $D=6$ dimensions.
For the first case the exact form of the four-point function of $\phi(x)$ has  been found \cite{Cardy:1985yy}; we shall use this result to test the accuracy of our method. $D=6$ is the upper critical dimensionality of the $\phi^3$ model, above which the classical mean-field value $\sigma=\frac12$ applies.

Our approach to conformal bootstrap only requires to know the spin and 
possibly other quantum numbers of the low-lying primary operators contributing
 to the OPE of $\phi(x)\phi(y)$. It is not necessary to know the detailed form of this OPE but simply its fusion rule that we write as
\beq
[\Delta_\phi]\times[\Delta_\phi]=\sum_i[\Delta_i,L_i],
\label{fusion}
\eeq
where $[\Delta,L]$  denotes the primary operator with scaling 
dimension $\Delta$ and spin $L$ and we set $[\Delta]=[\Delta,0]$.  
Note that this is a shorthand way of writing the conformal block expansion 
(\ref{bexpansion}) or even the first terms of a Clebsch-Gordan series 
of the conformal group $SO(D+1,1)$. 
\begin{table}[ht]
\centering
\begin{tabular}{c|c|c|c|c|c|}
\cline{2-5}
&\multicolumn{4}{c|}{2d Yang Lee model -- Truncated Fusion Rules}\\
\cline{2-6}
&\multicolumn{4}{c|}{
$[\Delta_\phi]\times[\Delta_\phi]=1+[\Delta_\phi]+[2,2]+[4]+[\Delta_4,4] $}&Exact\\
%\cline{2-5}
\protect\footnotesize 
&\protect\footnotesize 
  
&\protect\footnotesize 
 $+[\Delta_6,6]+[6,6] $ 
&\protect\footnotesize 
$+[\Delta_6,6]+[\Delta'] $
&\protect\footnotesize 
$+[6,6]+[\Delta'] $& results\\
\hline
\multicolumn{1}{ |c| }
  {$\Delta_\phi$}&-0.385(7) & -0.40033(1) & -0.40062(3) & -0.39777(2)&$-\frac25$ \\
\multicolumn{1}{ |c| }
  {$\Delta_4$}&3.70(4) & 3.5904(3) & 3.58961(5) & 3.5963(2) &$~~\frac{18}5$\\
\multicolumn{1}{ |c| }
 {$\Delta_6$}  & -- & 5.593(3) & 5.590(3) & --&$~~\frac{28}5$ \\
\multicolumn{1}{ |c| }
 {$\Delta'$}  & -- & -- & 7.60(1) & 7.61(8)&$~~\frac{38}5$ \\
\multicolumn{1}{ |c| } 
{$c$}   &-4.5(1) & -4.38(1) & -4.38(1) & -4.44(1)&$-\frac{22}5$ \\
\multicolumn{1}{ |c| }
{${\sf p}_{\phi}$} &-3.67(1) & -3.6524(3) 
& -3.6524(3) & -3.6557(3)&-3.65312.. \\
 \hline
\end{tabular}
\caption{The low-lying spectrum of primary operators of the 2d Yang-Lee model 
estimated for 4 different truncations of the fusion rule (\ref{YLfusion}) and compared with the exact results (last column). A rough estimate of the errors 
is obtained by the spread of the solutions generated by different set of equations. An example is depicted in figure \ref{fig:4}.}
\label{tab:1}
\end{table}
In a free scalar theory in $D$  
dimensions we have (see for instance (\ref{free3}))
 \beq
[\Delta_\phi]\times[\Delta_\phi]=1+[\Delta_{\phi^2}]+[D,2]+[D+2,4]+[D+4,6]+\dots,
\label{free}
\eeq
where $[D+L-2,L]$ is a conserved spinning operator. In an interacting
theory only the spin 2 -- the stress-tensor $T_{\mu\nu}$ -- is
generally conserved. Moreover in the $\phi^3$ theory $\phi^2$ is a
redundant operator, since at the non-trivial fixed point in $D<6$
dimensions it is proportional to $\partial^2\phi$ by the equation of
motion. Thus $\phi^2$ and its derivatives become descendant operators
of the only relevant operator $\phi$ of the Yang-Lee universality
class.  Thus the fusion rule that characterizes the low-lying spectrum
of this CFT is expected to be

\beq
[\Delta_\phi]\times[\Delta_\phi]=1+[\Delta_{\phi}]+[D,2]+[\Delta_4,4]+[\Delta']+\dots,
\label{YLfusion}
\eeq    
where  $[\Delta']$ in the perturbative renormalisation group analysis corresponds to $\phi^3$. This scalar cannot be neglected if we include in the fusion rule 
the spin 4 operator, since in $6-\epsilon$ dimensions 
$\Delta_{\phi^3}<\Delta_4$. Of course we expect also a term $[\Delta_6,6]$, 
like in a free-field theory, however its contribution becomes important only in 
the two-dimensional case.  We must also  add a scalar of dimension 4 corresponding to $T\bar{T}$.

 Comparison with the exact solution at $D=2$ shows that there are actually two primary operators of spin 6. One of them is conserved and coincides with $T^3$. 
Inserting truncations 
of such a fusion rule in \eq{det} we  found only isolated
solutions where the number  $M$ of fulfilled equations is larger than the number of unknowns 
$\Delta$'s, thus this CFT is easily truncable. To solve
  numerically these equations we used an iterative Newton–-Raphson
  method, which proves to be very effective\footnote{This method is implemented for instance in the {\tt FindRoot} function of Mathematica.}. In all the cases considered we worked with $M=7$ homogeneous equations. The number of unknown $\Delta$'s vary from 2 to 4 in the $D=2$ case (see table \ref{tab:1}), while for $D>2$ we have the 3 unknowns $\Delta_\phi,\Delta_4$ and $\Delta'$.

In table \ref{tab:1} we report the results of a truncation of this 
fusion rule to the first 4 terms and three different truncations of 6 terms. As expected the accuracy increases with the number of included conformal blocks. Inserting these 
$\Delta$'s in the linear system (\ref{inh}) and (\ref{homos}) we can
evaluate the OPE coefficients, in particular ${\sf p}_{\Delta_\phi}$
that turns out to be very close to the known exact result. Similarly
from ${\sf p}_{2,2}$ we can extract the central charge
$c=\Delta_\phi^2/{\sf p}_{2,2}$. 

We encountered isolated solutions also in $D>2$. The estimates for $D\le6$ 
generated by the truncated fusion rule (\ref{YLfusion}) are reported in tables
\ref{tab:2} and \ref{tab:3}. 
\begin{table}[ht]
\centering
\begin{tabular}{|c|c|c|c|c|c|}
\hline
\multicolumn{6}{|c|}{$D$ dimensional Yang-Lee model--the edge exponent $\sigma$} \\
\hline
$D$&bootstrap&Ising in $H$ &Fluids &Animals& $\epsilon-$expansion\\
\hline
2&-0.1664(5)&-0.1645(20)&-0.161(8)&-0.165(6)&(exact -1/6)\\
3&0.085(1)&0.077(2)&0.0877(25)&0.080(7)&0.079-0.091\\
4&0.2685(1)&0.258(5)&0.2648(15)&0.261(12)&0.262-0.266\\
5&0.4105(5)&0.401(9)&0.402(5)&0.40(2)&0.399-0.400\\
6&0.4999(1)&0.460(50)&0.465(35)&---&1/2\\
\hline
\end{tabular}
\caption{Estimates of $\sigma(D)$ from the truncated fusion rule 
(\ref{YLfusion}) compared with the best estimates  from strong coupling 
expansions \cite{Butera:2012tq}, fluids \cite{Lai:1995} and lattice animals
\cite{Hsu:2005}. A more complete list of estimates can be found in 
\cite{Butera:2012tq}. The bootstrap result for $D=2$ is calculated from tab \ref{tab:1}. }
\label{tab:2}
\end{table}
\begin{table}
\centering
\begin{tabular}{|c|c|c|c|}
\hline
\multicolumn{4}{|c|}{$D$ dimensional YL model-- $\Delta_4$ and $\Delta'$ } \\
\hline
$D$&${\sf p}_{\phi}$&$\Delta_4$&$\Delta'$\\
\hline
3&-3.88(1)&4.75(1)&5.0(1)\\
\hline
4&-2.72(1)&5.848(1)&6.8(1)\\
\hline
5&-0.95(2)&6.961(1)&6.4(1)\\
\hline
\end{tabular}
\caption{Estimates of $\Delta_4$ and $\Delta'$ and the OPE coefficient 
${\sf p_\phi}$ from the truncated sum rule (\ref{YLfusion}). These are, to the best of our knowledge,
the only known estimates for these quantities.} 
\label{tab:3}
\end{table}

\begin{figure}
\centering
\includegraphics[width=10 cm]{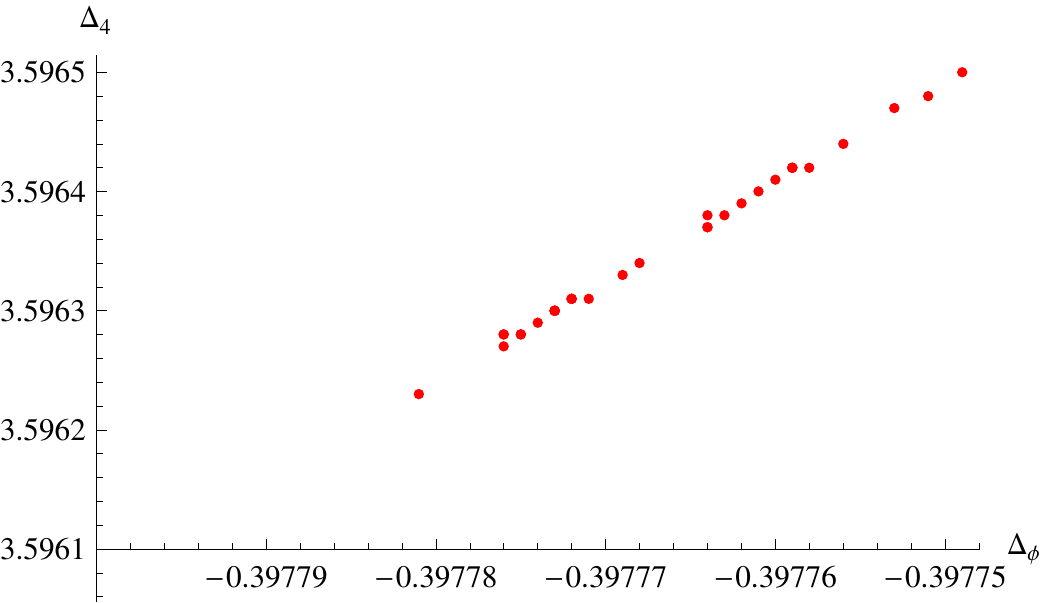}
\caption{Dispersion of the estimates of $\Delta_4$ versus
  $\Delta_\phi$ in the 2d Yang-Lee model. Notice the strong correlation between these two quantities.}
\label{fig:4}
\end{figure}

\begin{figure}
\centering
\includegraphics[width=12 cm]{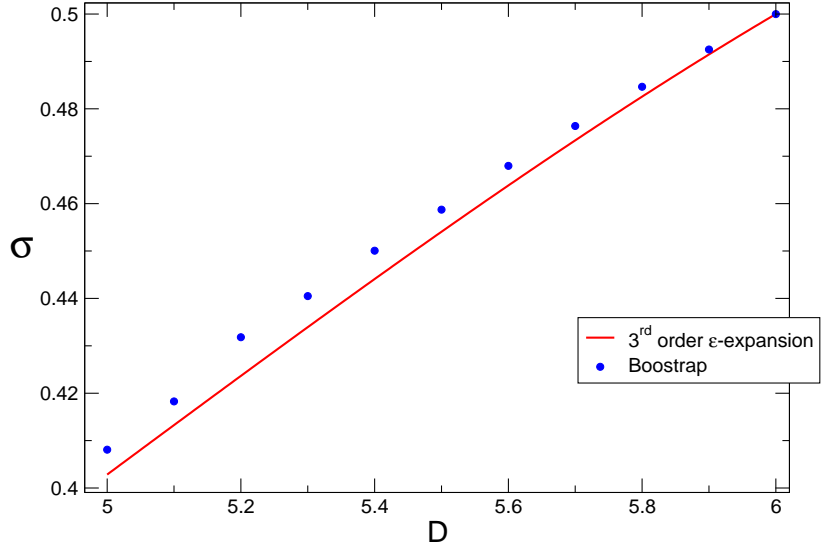}
\caption{Plot of the function $\sigma(D)$ given by \eq{YLepsilon} compared with the results of our bootstrap analysis for $5\le D\le6$. In this range the different resummations of the epsilon expansion do not give visually different results. The difference between the bootstrap estimates and the epsilon expansion are probably due to the contributions of higher order in $\epsilon$.}
\label{fig:5}
\end{figure}

\begin{figure}[ht]
\centering
\includegraphics[width=12 cm]{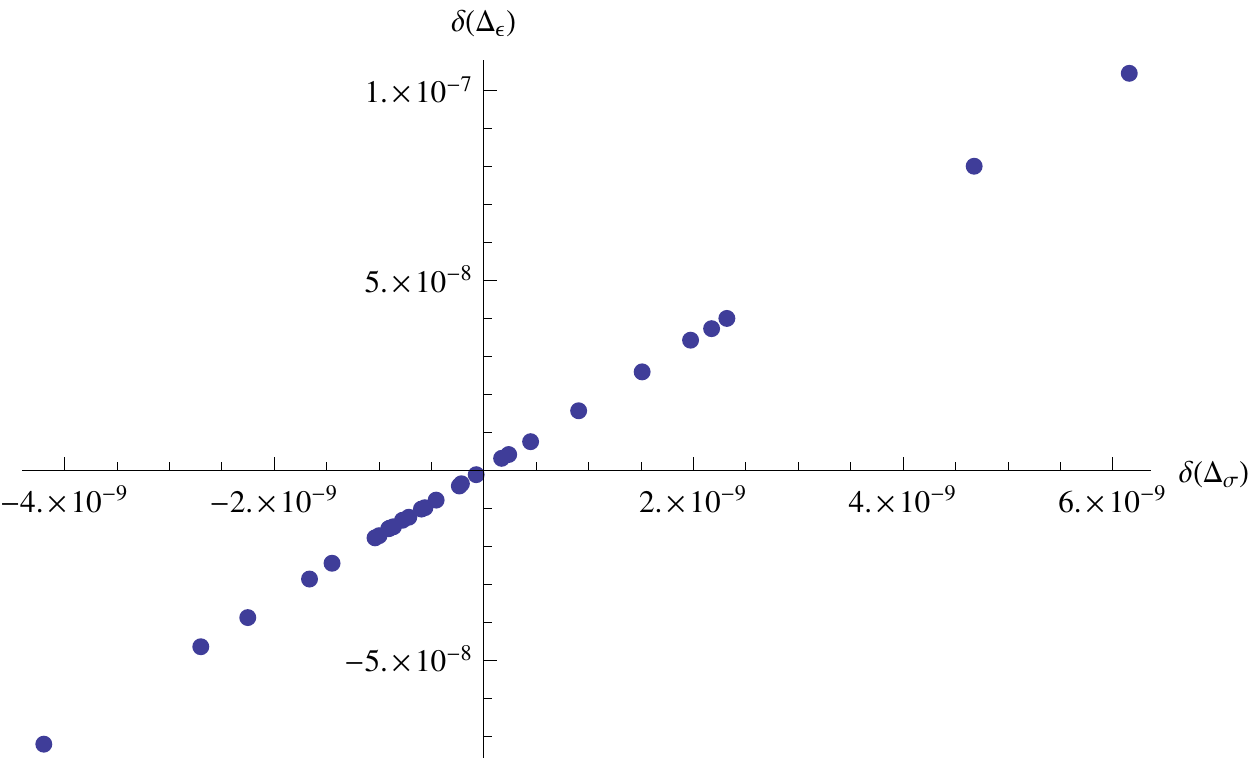}
\caption{Dispersion of the solutions in the plane $(\Delta_\sigma,\Delta_\varepsilon)$ at $\Delta_4=5.022$. The origin coincides with the mean value of these quantities. Notice the microscopic value of the spread.}
\label{fig:6}
\end{figure}

\begin{figure}[ht]
\centering
\includegraphics[width=12 cm]{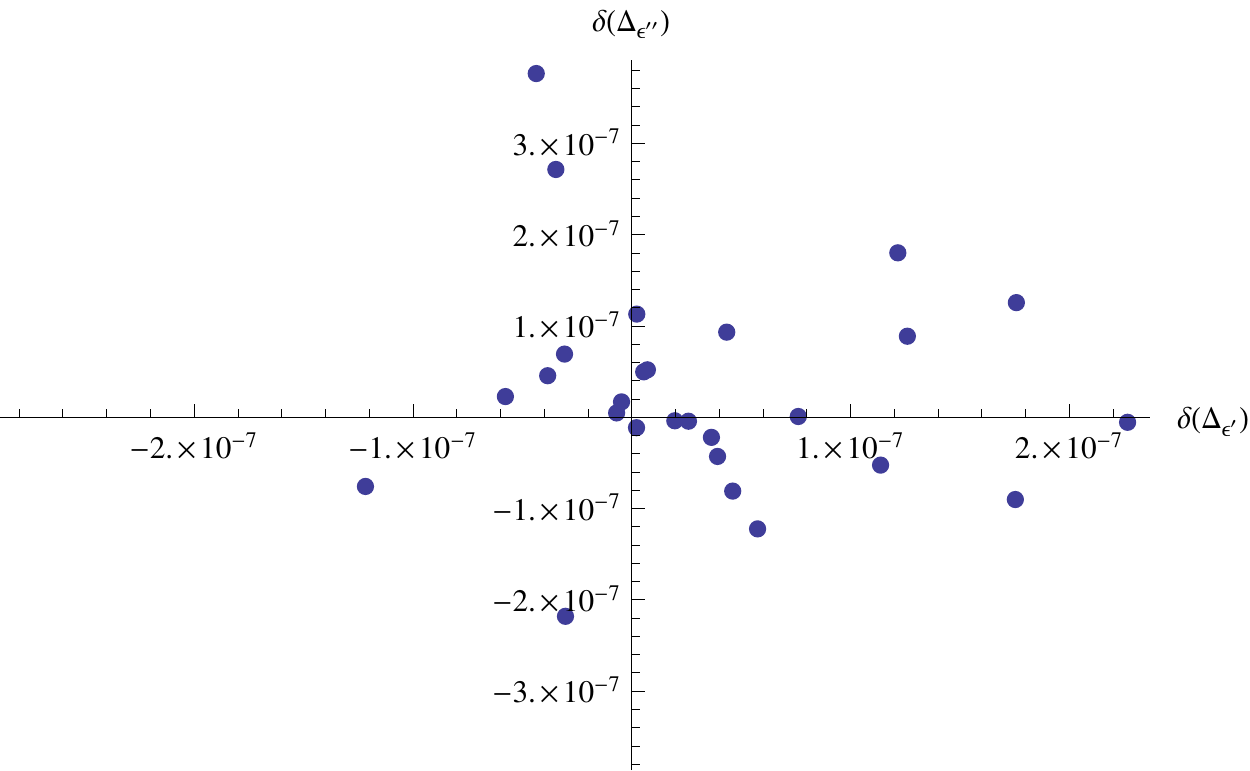}
\caption{Dispersion of the solutions in the plane $(\Delta_{\varepsilon'},
\Delta_{\varepsilon''})$ at $\Delta_4=5.022$. The origin coincides with the mean value of these quantities.}
\label{fig:7}
\end{figure}

\begin{figure}[ht]
\centering
\includegraphics[width=15 cm]{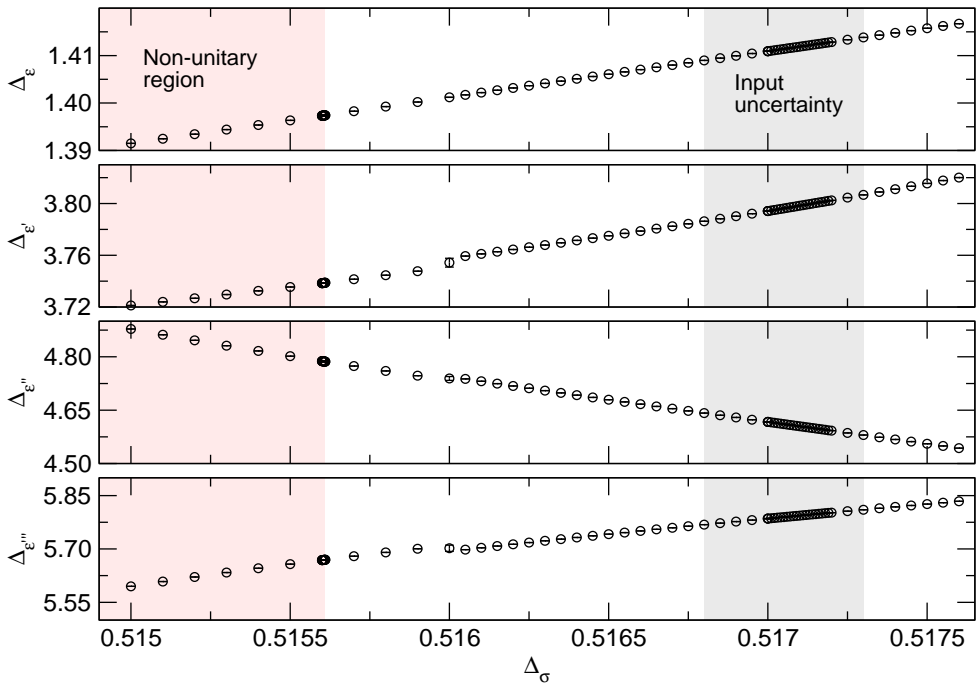}
\caption{Spectrum of the scalar operators of the one-dimensional family 
of solutions for the fusion rule  (\ref{fusion7}). 
The transition to the non-unitary region 
(the shadowed region on the left) is characterised by a change of sign
of the coupling ${\sf p}_{\epsilon''}$ (see
fig. \ref{fig:couplings}). 
Presumably, increasing the number of primaries in our analysis, the gap 
between the two shaded regions should shrink, in accordance with the conjecture that the critical Ising model saturates the unitarity 
bound \cite{ElShowk:2012ht}.
}
\label{fig:new}
\end{figure}

\begin{figure}[ht]
\centering
\includegraphics[width=15 cm]{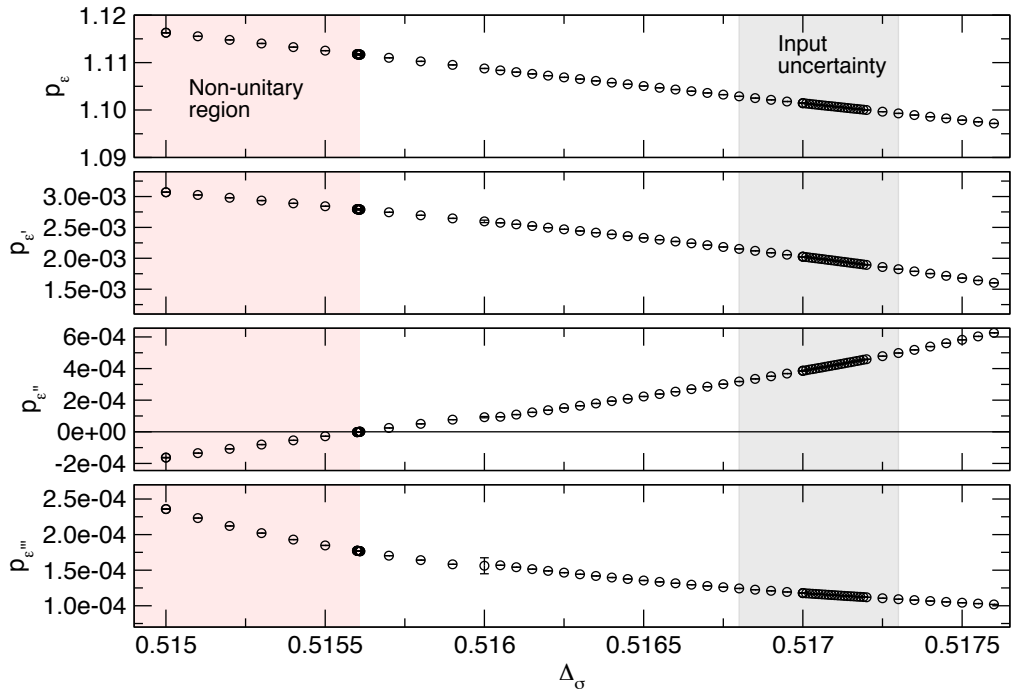}
\caption{Evolutions of the couplings of the scalar operators to the $\sigma\sigma$ channel 
for the fusion rule  (\ref{fusion7}) along the one-dimensional family of solutions. 
The transition to the non-unitary region 
(the shadowed region on the left) is characterised by a change of sign of the coupling ${\sf p}_{\epsilon''}$ . }
\label{fig:couplings}
\end{figure}

\begin{figure}[ht]
\centering
\includegraphics[width=10 cm]{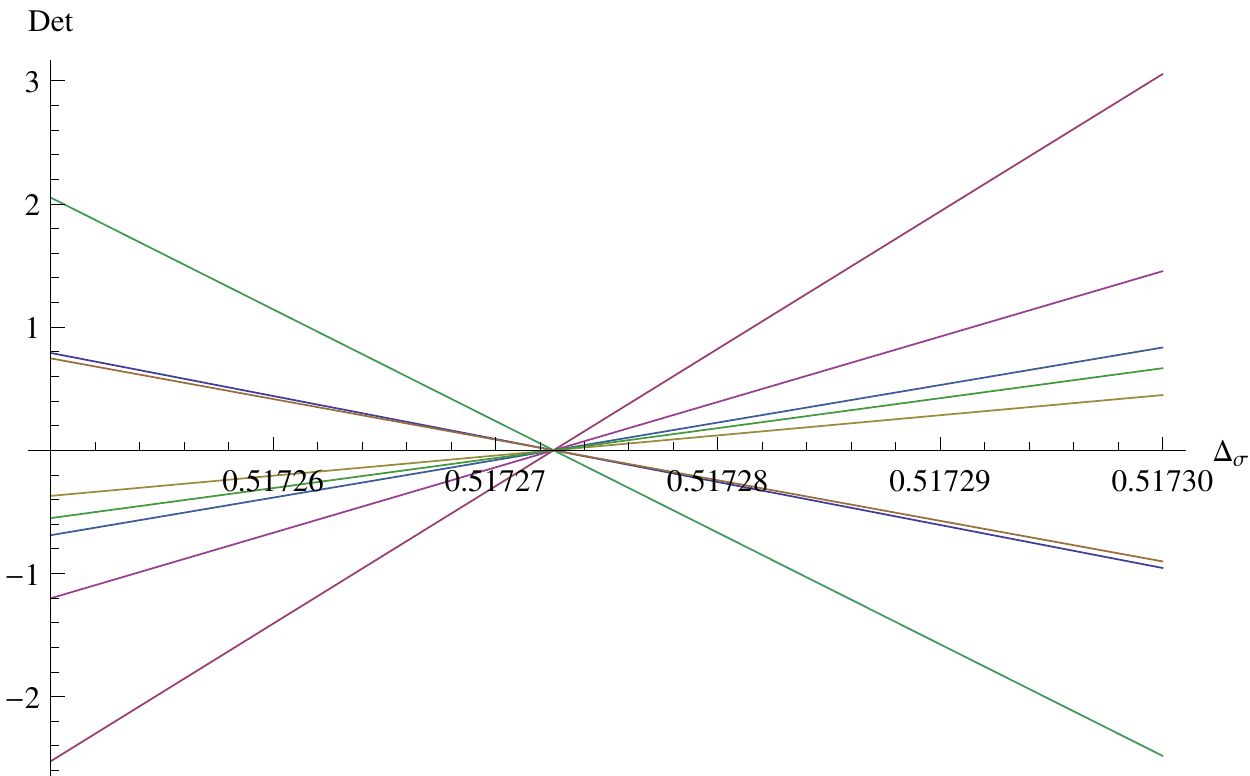}
\caption{The  zeros of the eight determinants associated to the solution 
(\ref{fusion7}) at $\Delta=\Delta_\sigma$, keeping all  other scaling dimensions fixed at the values of the exact solution. In view of the microscopic spread of the solutions shown in figure \ref{fig:6}, these different solutions are visually indistinguishable.}
\label{fig:8}
\end{figure}
\begin{figure}[ht]
\centering
\includegraphics[width=10 cm]{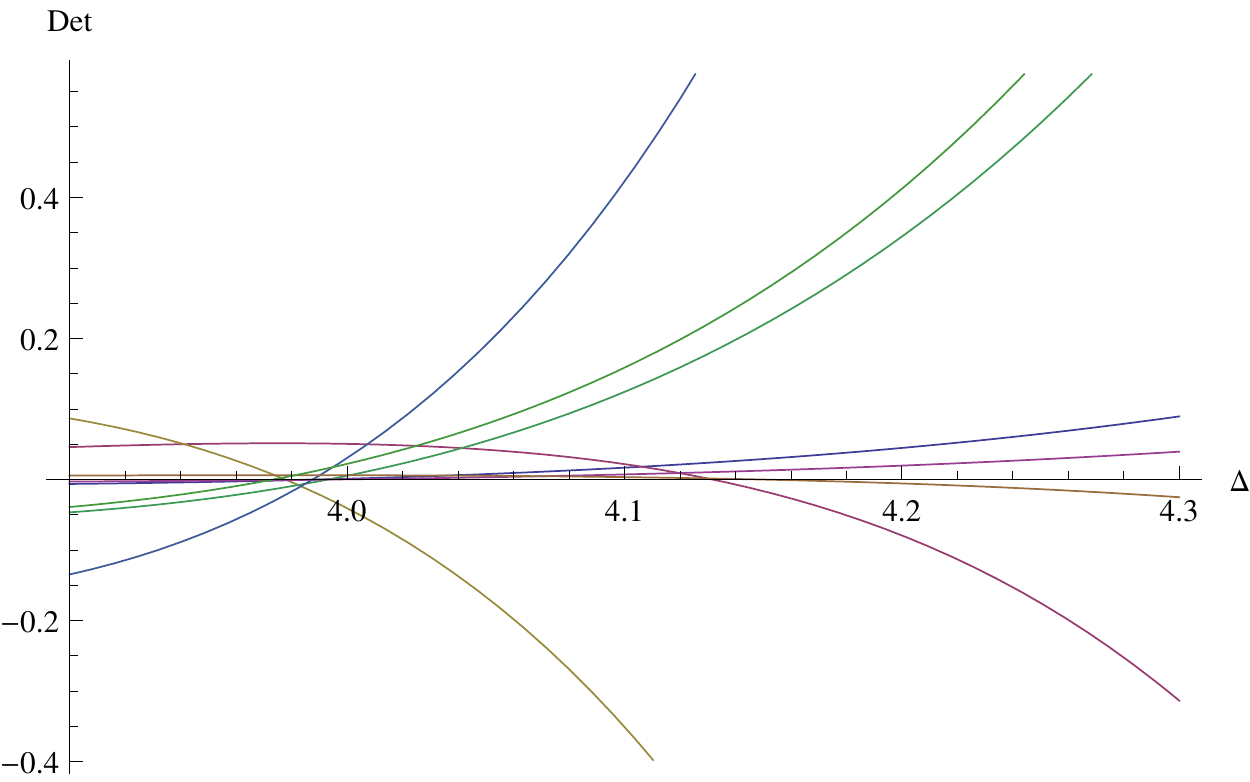}
\caption{Plot of the same determinants of figure \ref{fig:8} in the region
$\Delta\simeq 4$ were there is an approximate solution for $\Delta_{\sigma'}$. The spread of the zeros gives a rough estimate of the error.}
\label{fig:9}
\end{figure}
As we have already noticed, the present method works for any value of dimension. As such it is interesting to follow the flow of the spectrum from the six-dimensional free-field theory in $6-\epsilon$ dimensions. The epsilon expansion of 
$\sigma(D)$ known up to, and including, $\epsilon^3$ contributions is
\cite{deAlcantara:1980} 
\beq
\sigma(6-\epsilon)=\frac12 -\frac1{12}\epsilon-\frac{79}{3888}\epsilon^2+
\left(\frac{\zeta(3)}{81}-\frac{10445}{1250712}\right)\epsilon^3+O(\epsilon^4).
\label{YLepsilon}
\eeq

In figure \ref{fig:5} we plot this function in the range $5\le D\le6$ as well as 
the bootstrap estimates.

\section{3d critical Ising model}
\label{Ising}

We investigate the 3d critical Ising model with the same strategy applied to the Yang-Lee model, namely we start from the fusion rules of the free-field theory
in the upper critical dimension --  $D=4$ for a $\phi^4$ theory
-- and see how this is modified in the non-trivial fixed point 
in $4-\epsilon$ dimensions. 
This time the first redundant operator is $\phi^3$ which  does not 
contribute to the OPE of $\phi(x)\phi(y)$, being an odd operator,
so no further insight to simplify the fusion rule (\ref{free}) can be obtained . 
 If we include the spin 4 operator, 
we cannot neglect the leading irrelevant scalar primary $\varepsilon'$, corresponding to $\phi^4$, since $\Delta_{\phi^4}<\Delta_4$ in $4-\epsilon$ dimensions. 

A first solution of the bootstrap constraints (\ref{det}) is associated with the truncated fusion rule
\beq
[\Delta_\sigma]\times[\Delta_\sigma]=1+[\Delta_{\varepsilon}]+[\Delta_{\varepsilon'}]+
[3,2]+[\Delta_4,4],
\label{fusion4}
\eeq
with $\Delta_\sigma\equiv\Delta_\phi$. These scaling dimensions are related to the conventional  critical exponents by $\eta=2\Delta_\sigma-1$, 
$\nu=1/(3-\Delta_\varepsilon)$,  $\omega=\Delta_{\varepsilon'}-3$ and $\omega_{NR}=\Delta_4-3$, where $\eta$ is the magnetic exponent, $\nu$ the thermal exponent, 
$\omega$ is the exponent controlling the scaling corrections of
rotationally invariant operators and finally $\omega_{NR}$ is the one
controlling the correction to scaling of non-rotationally invariant
operators (see 
\cite{Pelissetto:2000ek}, Section 1.6.4).

 It turns out that  the bootstrap constraints (\ref{det}) applied to the 
$M=5$ homogeneous equations (\ref{homos}) with $m+n\le2$ are
fulfilled. 
In this case there is not an isolated solution like in Yang-Lee model,
but a one-parameter family of solutions. 
This result stems from the following:
the Newton--Raphson method requires an initial  
guess of the solution, then a sequence of approximate solutions is
generated rapidly converging to an accurate one.
 In the present case it turns out 
 that a small variation of the initial guess
yields a small change to the final point of the sequence, showing that
there is indeed a family of solutions. 
In order to verify that this family is one-dimensional  we kept  
the scaling dimension $\Delta$ of one of the unknowns as a free
external parameter
and we verified that for any given value of $\Delta$ the solution does not depend 
any longer on the initial guess. 
By varying the input parameter $\Delta$ we generated the one-dimensional family drawn in figures \ref{fig:new} and \ref{fig:couplings}.

Using as input $\Delta_4\equiv\omega_{NR}+3=5.0208(12)$, which is 
the most precisely known scaling dimension of the 3d critical Ising model 
(see \cite{Pelissetto:2000ek}, Section 3.2.1), we obtain a first rough estimate of the scaling dimensions of the operators involved in (\ref{fusion4}), namely
$\Delta_\sigma=0.5145(3)\,,\Delta_\varepsilon=1.3735(15)\,,\Delta_{\varepsilon'}=3.80(1)$.

A better approximation is obtained including also a spin 6 operator, 
and, since in the $\epsilon$ expansion the scaling dimensions of the
fields $\phi^6$ and $\phi^8$ are smaller than
$\Delta_6$, also these need to be included, resulting in the next to leading scalars  $\varepsilon'',\epsilon'''$.

Precisely we found that the truncated fusion rule
\beq
[\Delta_\sigma]\times[\Delta_\sigma]=1+[\Delta_{\varepsilon}]+
[\Delta_{\varepsilon'}]+[\Delta_{\varepsilon''}]+[\Delta_{\varepsilon'''}]+
[3,2]+[\Delta_4,4]+[\Delta_6,6],
\label{fusion7}
\eeq
admits an exact solution of the bootstrap constraints (\ref{det}) applied to the set of $M=8$ homogeneous equations (\ref{homos}) with $m+n\le3$ and $n<3$. 
The unknowns are the six quantities $\Delta_\sigma,\Delta_\varepsilon,
\Delta_{\varepsilon'},\Delta_{\varepsilon''}\Delta_{\varepsilon'''},\Delta_6$, so we can 
split the set of  homogeneous equations in 28 consistent subsystems 
and as many bootstrap constraints (\ref{det}) given by the vanishing
of determinants of $7\times7$ matrices. The corresponding spread of
the solutions is far smaller (see figures \ref{fig:6} and
\ref{fig:7}) than the error of the input parameter $\Delta_4$.  
In fig. \ref{fig:new} the one-parameter family of solutions is
depicted. Below $\Delta_\sigma=0.515607$ this line enters
 a non-unitary region characterised by a change of sign in ${\sf
  p}_{\varepsilon''}$. 
It is worth reporting that close to this family there are other one-parameter 
solutions which form an intricate and thick web in regions of the space of operator dimensions. Some of these lines intersect or bifurcate.
 Most of them correspond to non-unitary CFTs.

There is a number of consistency checks that can be exploited to enlarge the spectrum of estimated scaling dimensions. We expect in particular that the 
leading irrelevant odd  scalar $\sigma'$  fulfils the same fusion rule as $\sigma$, at least at this truncation level\footnote{Since $\phi^3$ is a redundant operator, this primary operator is 
associated with $\phi^5$ in the renormalisation group analysis in $4-\epsilon$ dimensions.}. This implies that if we treat $\Delta_\sigma$ as a free parameter
$\Delta$, keeping all the other scaling dimensions fixed, the 8
determinants which vanish  at $\Delta=\Delta_\sigma$ should also be
approximately zero at $\Delta=\Delta_{\sigma'}$, and this actually happens in figures \ref{fig:8} and \ref{fig:9}.   

Similarly, we expected that the fusion rule of the energy operator 
$\varepsilon(x)$ should coincide at this truncation level with that of 
$\sigma(x)$. However we did not find a corresponding solution. We found instead an approximate solution for the following enlarged fusion rule  
\beq
[\Delta_\varepsilon]\times[\Delta_\varepsilon]=1+[\Delta_{\varepsilon}]+
[\Delta_{\varepsilon'}]+[\Delta_{\varepsilon''}]+[\Delta_{\varepsilon'''}]+
[3,2]+[\Delta_2,2]+[\Delta_2',2]+[\Delta_4,4]+[\Delta_4',4]+[\Delta_6,6].
\label{fusion10}
\eeq
Here, besides all the operators of
(\ref{fusion7}),  contributions appear from three new spinning
operators, namely two recurrences of spin 2 and one recurrence of spin
4. In this case the conformal bootstrap constraints are associated
with determinants of $10\times10$ matrices, the number of homogeneous equations considered is $M=10$, while the number of unknown 
$\Delta$'s is 8 (remember that $\Delta_4$ is set as external
input). In order to have a sufficient number of homogeneous equations
we also considered the fusion rule (\ref{fusion10}) where in the LHS we replaced $\varepsilon$ with $\varepsilon'$. Unlike the solution of (\ref{fusion7}), where the accuracy of the zeros was of the order $10^{-16}$, here we found 
only  approximate zeros of the order $10^{-2}-10^{-3}$. It is important to notice that the estimate of 
scaling dimensions of the operators appearing in both fusion rules 
almost coincide. The resulting set of estimates for the whole set of primary 
operators considered in our analysis is reported in tables \ref{tab:4} and
\ref{tab:5}. Finally, inserting these values in the linear systems (\ref{inh}) 
and (\ref{homo}) we can extract the corresponding OPE
coefficients. They are reported in figure \ref{fig:couplings} and in table \ref{tab:6}. A word of
caution: our data are affected by an unknown systematic
error due to the truncation in the number of operators. Such
systematic error is bound to decrease when an higher number of operators is
used. The errors reported in the tables only take into account the
spread of the solutions and the uncertainty of the input parameter.

\begin{table}[ht]
\centering
\begin{tabular}{|c|c|c|c|c|c|c|}
\hline
scalar operators&$\sigma$&$\sigma'$&$\varepsilon$&$\varepsilon'$&$\varepsilon''$&$\varepsilon'''$\\
\hline
\hline
$\Delta$, best estimates&0.51813(5)&$\gtrsim4.5$&1.41275(25)&3.84(4)&4.67(11)&--\\
\hline
$\Delta$, bootstrap&0.51705(25)&4.05(5)&1.4114(24)&3.796(10)&4.61(3)&5.79(2)\\
\hline
\end{tabular}
\caption{ Low-lying scalar primary operators and their scaling
  dimensions obtained by the solution of the conformal bootstrap
  constraints (\ref{det}). 
The comparison values are taken from \cite{Martin2010} and 
\cite{Pelissetto:2000ek}. Recent calculations in the 
functional renormalisation group approach (see \cite{Litim:2003kf} and 
references therein) predict  much higher values for 
$\Delta_{\varepsilon''}$ and $\Delta_{\varepsilon'''}$( see Tab. \ref{tab:7}). However they also 
give $\eta=0$, implying that in their case $\sigma(x)$ is a free field.  }
\label{tab:4}
\end{table}

\begin{table}[ht]
\centering
\begin{tabular}{|c|c|c|c|c|c|}
\hline
spinning operators&~~$[\Delta_2,2]$&$[\Delta_2',2]$&~~~~$ [\Delta_4,4]  
$~~\,~~&
$[\Delta_4',4]$&$[\Delta_6,6]$ \\
\hline
\hline
$\Delta$, best estimates&--&--&5.0208(12)&--&7.028(8)\\
\hline
$\Delta$, bootstrap&5.117(1) &6.20(1)&input&6.70(1)&7.065(3)\\
\hline
\end{tabular}
\caption{Low-lying  primary operators with spin and their scaling dimensions 
obtained by the solution of the conformal bootstrap constraints (\ref{det}).
 The input value for $\Delta_4$ is taken from 
\cite{Pelissetto:2000ek}.  The estimate for the spin 6 operator is extracted from \cite{Komargodski:2012ek}. }
\label{tab:5}
\end{table}
\begin{table}[ht]
\centering
\begin{tabular}{|c|c|c|c|c|c|c|}
\hline
${\sf p}_{\varepsilon}$&${\sf p}_{3,2}$&${\sf p}_{\varepsilon'}$&
${\sf p}_{\varepsilon''}$&${\sf p}_{\varepsilon'''}$&${\sf p}_{\Delta_4}$&${\sf p}_{\Delta_6}$\\
\hline
1.101(2)& 0.2855(4)& 0.0019(2)& 0.00041(1)&
 0.000116(7)&  0.01612(2)& 0.001541(2)\\ 
\hline
 1.1117 & 0.28326&  0.0027906& 0.4(1.3)E-8& 0.0001768& 
0.01601& 0.0015300  \\
\hline
\end{tabular}
\caption{In the first row we report the values of the couplings related to the
results presented in tables \ref{tab:4} and
\ref{tab:5} for the fusion rule (\ref{fusion7}). The error originates from the uncertainty on the input parameter. 
In the second row we report the values of the couplings for the point $\Delta_\sigma=0.515607$, which lies at the boundary of the
 unitarity region (see fig. \ref{fig:new}). 
The error due to the spread of the solutions is negligible (it ranges
from $10^{-13}$ to $10^{-8}$) hence it is not reported, except for 
${\sf p}_{\varepsilon''}$.
The central charge extracted from the first row is
$c/c_{free}=0.9364(22)$; while the second row we obtain 
$c/c_{free}=0.93853(1)$. 
Presumably, increasing the number of included 
primaries in our analysis, the gap between the point at the boundary and the Ising point should shrink,
in accordance with the conjecture that the critical Ising model saturates the unitarity 
bound \cite{ElShowk:2012ht}.}
\label{tab:6}
\end{table}

It is worth noting that, beside contributing to the to the
  $n$-point functions, the scalar operators also play an important role in  
the description of the thermodynamic functions at criticality. 
For instance the critical behaviour of the magnetic susceptibility in terms of the reduced temperature $t$ is given by the Wegner expansion
$$\chi=t^{-\gamma}(a_{0,0}+a_{0,1}t+a_{0,2}t^2+\dots+a_{1,1}t^{\omega\nu}+
a_{1,2}t^{2\omega\nu}+\dots+a_{2,1}t^{\omega_2\nu}+\dots+
a_{3,1}t^{\omega_3\nu}+\dots),
$$
where the leading exponent $\gamma$ is not  directly evaluated  in the 
bootstrap approach, but can be expressed in terms of $\nu$ and $\eta$ through
$\gamma=\nu(2-\eta)$. The leading and subleading correction-to-scaling 
exponents are related to the dimensions of the scalar operators by 
$\Delta_{\varepsilon'}=3+\omega$, $\Delta_{\varepsilon''}=3+\omega_2$ and  
$\Delta_{\varepsilon'''}=3+\omega_3$. The determinations of these exponents in our work are compared with other determinations in Table \ref{tab:7}.

\begin{table}[ht]
\centering
\begin{tabular}{|c|c|c|c|c|c|}
\hline
method&~~$\eta$&$\nu$&~~~~$ \omega  
$~~\,~~&
$\omega_2$&$\omega_3$ \\
\hline
\hline
 this work &0.0341(5) &0.629(1)&0.80(1)&1.61(3)&2.79(2)\\
\hline
HT \cite{phi4pisa} &0.03639(15)&0.63012(16)&0.83(5)&--&--\\
\hline
MC \cite{Martin2010} &0.03627(10)&0.63002(10)&0.832(6)&--&--\\
\hline
RG \cite{Litim:2010tt,Litim:2003kf} &0.034(5)&0.630(5)&0.82(4)&2.8--3.7&5.2--7.0\\
\hline
SF \cite{Newman:1984zz} &0.040(7)&0.626(9)&0.855(70)&1.67(11)&--\\
\hline
UB \cite{El-Showk:2014dwa}&0.03631(3)&0.62999(5)&0.8303(18)&4&7.5\\
\hline
\end{tabular}
\caption{ Leading exponents and leading and subleading correction-to-scaling exponents  obtained using high 
temperature (HT) expansions,  Monte Carlo (MC) calculations,
functional renormalization group (RG), scaling field (SF) methods and unitarity bounds (UB) on solutions of crossing simmetry compared 
with the results of this work.}
\label{tab:7}
\end{table}

In conclusion, in this paper we have developed a general method implementing the crossing symmetry constraints in a large class of CFT, both unitary 
and not. Starting with the knowledge of the fusion rules, this method can
generate systematically the correlation functions, by searching for the zeros 
of certain $N\times N$ determinants made up of multiple derivatives of 
generalized hypergeometric functions. Such a study can be performed
with the use of modest computing power, indeed all the results here presented have been
obtained by using a single workstation. The method's application to the Yang -Lee edge 
singularity for  $2\le D\le6$ and to the 3d critical Ising model gives rather accurate results and can be further improved by enlarging the number $N$ of primary operators included in the analysis.
                 
\hskip 1 cm

{\sl Added Note}

After this work was completed, we became aware of the preprint \cite{El-Showk:2014dwa}
 where the low-lying spectrum of the 3d critical Ising model is precisely calculated under the assumption that this model has the smallest central 
charge, by analogy with the two-dimensional case. In this way very accurate values of the scaling dimensions $\Delta_\sigma$ and $\Delta_\varepsilon$ are obtained, which are more precise than our estimates with the 
determinant method.

In the quoted paper strong numerical evidence is reported, showing that some operators disappear from the spectrum as one approaches the putative 3d Ising 
point.  Our method, which is able to follow our one-parameter family of solutions also in the non-unitary region
  (see fig. \ref{fig:new}),  provides a simple explanation of this
phenomenon.
Moving across the two regions at least one of our couplings changes
sign. It follows that the corresponding operator decouples at the boundary
of the unitarity region, where the putative critical Ising point
should be.
In our solution we can see the vanishing of the coupling of
one operator (see tab. \ref{tab:6}) and presumably, enlarging the number
of primaries included in the analysis, the number of decoupled
operators at the boundary should increase. 
The operator that decouples is the scalar at $\Delta\simeq 4.60$ that we
identify with $\varepsilon''$. 
Despite this decoupling, this operator does not disappear from the
spectrum of the CFT at the boundary since its coupling to the
$\varepsilon\varepsilon$ channel is still different from zero. 
This operator is not present in the analysis of
\cite{El-Showk:2014dwa}. 
A possible explanation for such a discrepancy could be that the coupling of this operator 
is rather small along the entire one-parameter family of solutions (see fig \ref{fig:couplings}), and the method of  \cite{El-Showk:2014dwa} cannot detect operators with too 
small couplings.  

\section*{Acknowledgements}
We would like to thank S. Rychkov for useful discussions and insights.

\end{document}